\documentclass{article}[12pt]
\usepackage[utf8]{inputenc}
\usepackage[russian]{babel}
\usepackage{graphicx}
\renewcommand{\baselinestretch}{1.3}
\begin{document}
\begin{center}

Astrophysics, Vol. 58, No. 4, December, 2015

\bigskip

0571-7256/15/5804-0453 (C)2015 Springer Science+Business Media New York
Original article submitted August 18,  2015.  Translated from Astrofiz ika, Vol. 58, No. 4, pp. 487-504
(November 2015).

\bigskip

{\bf \Large EXTENDING  THE $H\alpha$ SURVEY FOR THE  LOCAL VOLUME GALAXIES }

\bigskip

{\large I. D. Karachentsev, S. S. Kaisin, and E. I. Kaisina}

\bigskip

{\em Special Astrophysical Observatory of the Russian Academy of Sciences;  e-mail: ikar@sao.ru}

ABSTRACT

\end{center}

{\em Images in the $H\alpha$ emission line are presented for 35 nearby objects observed with the 6-m BTA telescope.
Three of them, NGC~3377, NGC~3384, and NGC~3390, are bright E and S0 galaxies, one is an edge-on  Sd galaxy
UGC~7321,  two are remote globular clusters associated with  M~31, and the rest are dwarf galaxies 
 of morphological types dIr, dTr, dSph, BCD, and Sm.  The measured  $H\alpha$ fluxes are used to estimate
the integral $(SFR)$ and specific $(sSFR)$ star formation rates for these galaxies.  The values of $\log[sSFR]$
for all these objects lie below a limit of $-0.4$(Gyr$^{-1})$.  We note that the emission disk for the nearest superthin 
edge-on galaxy UGC 7321 has an extremely large axis ratio of $a/b = 38.$ }

\bigskip


{\section{Introduction.}

Since 2004 a program for surveying nearby galaxies in the Balmer $H\alpha$  line has been under way with the 6-m
 telescope at the Special Astrophysical Observatory of the Russian Academy of Sciences.  The purpose of this
program is to determine the star formation rate in the galaxies lying within a fixed local volume by measuring their
integral flux in the  $H\alpha$ line.

At the beginning of this survey in 2004, the ''Catalog of Nearby Galaxies,'' CNG [1], included 450 galaxies
with distances $D < 10$ Mpc.  Over the last 10 years, this sample has increased by a factor of two as a result of 
a wide 
field survey of the sky in the visible range and in the 21-cm neutral hydrogen line.  The latest version of the ''Updated
Nearby Galaxy Catalog'',  UNGC [2],  contains 869 objects with estimated distances  $D < 11$ Mpc and is still being
extended in the form of a data base [3] that can  be found on the internet at  http://www.sao.ru/lv/lvgdb.

Thus far we have obtained images in the  $H\alpha$ line for $\sim$400 nearby galaxies in the northern sky [4-11 ].  In order
to avoid undesirable selection effects we have observed galaxies of all morphological  types and luminosities without
exception.  These $H\alpha$-images of the galaxies and their $H\alpha$ fluxes, which are given in the data base [3], provide a
unique opportunity for research on the distribution of star formation regions in galaxies of different types with high
spatial resolution.  We have published a survey of the basic characteristics of star formation in nearby galaxies [12]
with data on measurements  of the $H\alpha$  fluxes for $\sim$200 galaxies in the Local volume  taken into account by a program
analogous to that initiated by Kennicutt [13].

Detailed charts of the distribution of ionized hydrogen in nearby galaxies together with data on  the distribution of
young blue stars in these galaxies obtained by the GALEX survey [14] make it possible to establish the
star formation rate  on time scales ranging from $\sim$10 to  $\sim$100 million years.  Multiaperture observations of these
galaxies in the $HI$ line [15,16] can be used to study the distribution and kinematics of neutral hydrogen and, thereby,
to gain a better understanding of the mechanism for conversion of the gas into stars.  The massive $H\alpha$-survey of nearby
galaxies also serves as a basis for selection and subsequent analysis of the kinematics of the most active objects with
the aid of a Fabry-Perot etalon [17,18].

Here we present $H\alpha$-images, values of the  $H\alpha$-flux, and estimates of the star formation rate $(SFR)$ for 35 objects
in the Local volume.  Of these objects, 30 are dwarf galaxies, 3 are massive E, S0 galaxies, and 2 are globular clusters in the 
remote periphery of M~31.

\section{Observations and data processing.}

The galaxies were observed in the primary focus of the BTA telescope with the SCORPIO focal reducer
equipped with a  $2048\times2048$ pixel CCD array over the period from October 2014 through March 2015.  The optical
system provided a field of view of $\sim6^{\prime}$ with a scale of $0.185^{\prime\prime}$/pixel. An interference $H\alpha$
filter with a width
 FWHM=75\AA \,and an effective wavelength of 6555\AA \, was used for the observations.  Images in the continuum were
taken with middle-band centered SED607 ($\Delta\lambda=167$\AA,
 $\lambda_e=6063$\AA) and SED707 ($\Delta\lambda=207$\AA, $\lambda_e=7036$\AA) filters .
 All the
objects were observed with the same set of filters, since the galaxies had a small range of  radial velocities.  The total
exposure was about 20  min for each object.

The observational data were processed by a set of standard procedures that included dark frame subtraction, then
the image was normalized to a flat field obtained at twilight and traces of cosmic
particles were eliminated by comparing images and sky background substracted.
 The continuum images were
normalized to the image in the $H\alpha$-filter using 10--30 stars and then subtracting. The $H\alpha$-flux of a  galaxy was
determined from the $H\alpha$-image after background subtraction. The flux was calibrated using images of spectrophotometric
 standard stars observed on the same nights. The typical error in the  measurement of the logarithm of the
$H\alpha$-flux was determined mainly by the weather conditions and was $\sim$0.1 dex.

\section{Observational results.}

A mosaic of 35 pairs of observed images of galaxies is shown in Fig. 1.  The left images of each  pair correspond to
combined exposure in the $H\alpha$ and the continuum, and the right  images, to the image in  $H\alpha$ with the continuum
 subtracted.
The name of each object, the image scale, and the ''north and east'' orientations are indicated on the right images.
Bright stars, as well as stars and galaxies with an anomalous color, show up in the right images as residual ''stump''.
Another reason for imperfection of the subtracted images was variable atmospheric turbulence.  A faint interference
pattern can be seen in some of the images owing to deficiencies in the procedure  for flat field division.

We corrected the measured ranges (in units of erg/cm$^2$/s) of the $H\alpha$-fluxes,  $F(H\alpha$), of the galaxies for Galactic
absorption in accordance with [19].  The corrected value $F_c(H\alpha$) served as an estimate of the integral star formation
rate of a galaxy [20] (in units of   $M_{\odot}/$year):

$$\log [SFR] =\log F_c(H\alpha) + 2 \log D +8.98,\eqno(1)$$
where $D$ is the distance to a galaxy in Mpc.  Here we neglect the contribution to the flux from the [$NII$] emission
doublet, as well from internal absorption in the galaxy, since both effects are small for dwarf galaxies [21,13], which
are a majority of the observed objects.  The only exception was the spiral galaxy UGC~7321, which is seen edge on 
and has an internal absorption of 0.66$^m$ in the $H\alpha$-line [21].

The main parameters of the observed galaxies are summarized in Table 1, the columns of which list the
following: (1) the name of the galaxy; (2) the equatorial coordinates at epoch 2000.0; (3) the integral $B$-magnitude; (4)
the morphological type according to data from the UNGC catalog [2]; (5) the distance in  Mpc [2]; (6) the method
by which this distance was determined ( {\em rgb} \ from the luminosity of stars in the red giant branch;  {\em sbf} \ from surface
brightness fluctuations;  {\em TF} \ from the Tully-Fisher relation between the rotation amplitude of a galaxy and its
luminosity;  {\em mem} \ from membership in known groups;  {\em h} \ from the radial velocity relative to the centroid of the Local
group for a Hubble parameter $H_0=73$ km/s/Mpc; and, {\em txt} \ from the texture of the image of a dwarf galaxy (this estimate
is extremely uncertain)); (7) the logarithm of the $H\alpha$-flux we have measured; (8) the logarithm of the integral star
formation rate according to the formula given above; (9) the integral star formation rate of the galaxy,

 $$\log[SFR(FUV)]=\log F_c(FUV) + 2 \log D -6.78, \eqno(2)$$
determined from its far ultraviolet flux ($\lambda_e = 1539$\AA, FWHM = 269\AA) measured by the GALEX satellite [14] with
a correction for absorption of light in our Galaxy; and, (10) the specific star formation rate normalized to unit stellar
mass of the galaxy.  Several estimates of the $H\alpha$-flux obtained through cirrus are  indicated by a colon.

   \renewcommand{\baselinestretch}{0.8}
\begin{table}
   \caption{Main Properties of the Observed Galaxies.} 
   \begin{tabular}{lcclrlrrrr} \hline
  Name     &  RA (2000.0) Dec&   $B_t$&    T&    D &  method&  log F(H$\alpha)$ & log SFR & log SFR& log sSFR \\ 
	   &                 &     &     &      &      &         & (H$\alpha$) &   (FUV) &  (H$\alpha$)\\ 

	   &                 &  mag&     &  Mpc &      &erg/cm$^2$s&  $M_{\odot}$/yr& M$_{\odot}$/yr&  yr$^{-1}$\\ \hline

  (1)      &        (2)      & (3) &  (4)&  (5) & (6)  &  (7)    &  (8) &   (9)   &  (10)\\ \hline
            & & & & & & & & &\\
PAndAS-04  & 000442.9+472142 & 18.4&  g.c&  0.78& mem  &$<-$15.71  &$<-$6.82& $ -$5.67  &$<-$12.42 \\
PAndAS-56  & 012303.5+415511 & 17.6&  g.c&  0.78& mem  &$<-$14.53  &$<-$5.72& $<-$6.47  &$<-$11.50 \\
 $[$TT2009$]$25 & 022112.4+422151 & 17.9&  dTr&  9.8 & mem  &$<-$15.41  &$<-$4.39& $ -$4.16  &$<-$11.58 \\
 $[$TT2009$]$30 & 022254.7+424245 & 18.9&  dIr&  9.8 & mem  &$<-$15.29  &$<-$4.27& $ -$4.25  &$<-$11.07 \\
KKH 30     & 051742.0+813727 & 17.5&  dTr&  9.0 & txt  &$<-$15.12  &$<-$4.17& $ -$3.53  &$<-$11.46 \\
	   &                 &     &     &      &      &         &      &         &        \\
N2146sat   & 062413.2+775753 & 17.5&  dTr&  9.0 & txt  &$<-$15.11  &$<-$4.16& $ -$3.82  &$<-$11.45 \\
AGC188955  & 082137.0+041901 & 17.5&  dIr& 14.5 & TF   &$ -$13.43  &$ -$2.11& $ -$2.24  &$  -$9.73 \\
LV0853+3318& 085326.8+331819 & 19.5&  dIr&  9.4 & mem  &$ -$14.15  &$ -$3.20& $ -$3.60  &$  -$9.66 \\
LV0926+3343& 092609.4+334304 & 17.8&  Sm & 10.3 & TFb  &$ -$13.76  &$ -$2.74& $ -$2.39  &$ -$10.04 \\
LV1017+2922& 101726.5+292211 & 16.6&  BCD&  5.4 & h    &$ -$13.97  &$ -$3.50& $   -$    &$ -$10.64 \\
	   &                 &     &     &      &      &         &      &         &        \\
AGC749315  & 102906.4+265438 & 19.1&  BCD& 11.0 & h'   &$ -$15.00  &$ -$3.91& $ -$3.04  &$ -$10.66 \\
AGC200499  & 103808.0+102251 & 14.4&  BCD& 13.9 & h    &$ -$13.05  &$ -$1.73& $   -$    &$ -$10.72 \\
NGC3377    & 104742.4+135908 & 11.2&  E  & 10.9 & sbf  &$ -$12.84  &$ -$1.75& $ -$2.25  &$ -$12.18 \\
NGC3384    & 104816.9+123745 & 10.9&  S0 & 11.4 & sbf  &$<-$15.32  &$<-$4.20& $   -  $  &$<-$14.94 \\
NGC3413    & 105120.7+324558 & 13.2&  Sm & 12.0 & TF   &$ -$12.13  &$ -$0.92&  $-$1.04  &$ -$10.27 \\
	   &                 &     &     &      &      &         &      &         &        \\
KKH 73     & 115006.4+554700 & 17.3&  dTr&  9.0 & txt  &$<-$15.12  &$<-$4.23& $<-$4.48  &$<-$11.49 \\
NGC3990    & 115735.6+552731 & 13.4&  S0 & 10.3 & sbf  &$<-$15.12  &$<-$4.10& $   - $   &$<-$13.63 \\
LV1157+5638& 115754.2+563816 & 17.1&  dIr&  7.0 & h    &$ -$13.0: &$ -$2.3:&$ -$2.86  &$  -$9.5: \\
UGC7257    & 121503.0+355731 & 14.4&  Sm &  8.8 & TF   &$ -$12.22  &$ -$1.32& $ -$1.31  &$  -$9.63 \\
UGC7321    & 121734.0+223225 & 14.1&  Sd & 17.2 & TF   &$ -$12.71  &$ -$0.97& $ -$0.23  &$ -$10.54 \\
	   &                 &     &     &      &      &         &      &         &        \\
LV1218+4655& 121811.1+465501 & 16.8&  Sm &  6.5 & h    &$ -$14.2: &$ -$3.6:& $-$2.89  &$ -$10.8: \\
LV1219+4718& 121927.2+471845 & 18.0&  dIr&  7.8 & mem  &$<-$15.19  &$<-$4.41&  $-$3.81  &$<-$11.28  \\
KK136      & 122040.6+470003 & 17.5&  dSp&h 7.8 & mem  &$<-$15.08  &$<-$4.30&  $  - $   &$<-$12.08 \\
KUG1218+387& 122054.9+382549 & 15.4&  BCD&  8.0 & h    &$ -$13.52  &$ -$2.71&  $-$2.12  &$ -$10.65 \\
UGC7427    & 122155.0+350305 & 15.9&  dIr&  9.7 & TF   &$ -$13.13  &$ -$2.17&  $  -$    &$ -$10.06 \\
	   &                 &     &     &      &      &         &      &         &        \\
DDO123     & 122608.1+581921 & 14.5&  Sm & 10.5 & TF   & $-$12.8: & $-$1.8:& $  - $   &$ -$10.4: \\
SBS1224+533& 122652.6+530619 & 16.1&  BCD&  5.4 & h    & $-$13.4: &$ -$3.0:& $-$2.65  &$ -$10.3: \\
DDO131     & 123158.6+294235 & 15.3&  dIr&  8.1 & mem  & $-$12.91  &$ -$2.09&  $-$1.94  &$ -$10.09 \\
NGC4509    & 123306.8+320530 & 14.1&  BCD& 10.1 & TF   & $-$12.20  &$ -$1.20&  $-$1.26  &$  -$9.94 \\
IC3583     & 123643.5+131534 & 13.3&  Im &  9.5 & rgb  & $-$12.62  &$ -$1.63&  $-$1.24  &$ -$10.44 \\
	   &                 &     &     &      &      &         &      &         &        \\
UGCA294    & 124438.1+282821 & 14.8&  BCD&  9.9 & TF   & $-$12.7: &$ -$1.7:& $-$1.57  &$ -$10.1: \\
IC3840     & 125146.1+214407 & 16.9&  dIr&  5.5 & TF   & $-$13.60  &$ -$3.10&  $-$2.69  &$ -$10.15 \\
DDO169NW   & 131520.1+473237 & 18.0&  dIr&  4.2 & mem  & $-$13.84  &$ -$3.60&  $-$3.02  &$  -$9.92 \\
CGCG189-050& 131704.9+375708 & 15.6&  BCD&  5.0 & h    & $-$12.78  &$ -$2.39&  $  - $   &$  -$9.84 \\
AF7448-001 & 225935.3+164611 & 17.2&  dIr&  5.0 & TF   & $-$13.63  &$ -$3.16&  $-$3.24  &$ -$10.11 \\
\hline
\end{tabular}
\end{table}

\section{Distinctive features of the observed objects.}

  {\bf PAndAS-04, PAndAS-56}. Two globular clusters at the distant periphery of M~31 from the catalog of [22]
with projected distances from the center of M~31 of 124 and 103 kpc, respectively.  Such isolated objects may be centers
to which intergalactic gas accretes.  Data from GALEX [14] show that PAndAS-04 has a significant flux in the far
ultraviolet.  But, according to our measurements, the $H\alpha$-flux from both globular clusters lies below  the detection
threshold.

{\bf [TT2009]25, [TT2009]30}.  Two candidates for satellites of the massive spiral galaxy NGC~891 detected in
[23].  A recent measurement of the radial velocity of the brighter dwarf object [TT2009]25 confirms its physical
coupling to the galaxy NGC~891 [24].  Both dwarf galaxies have been detected as faint FUV sources in the GALEX
survey, but their $H\alpha$-fluxes were below the detection threshold in our observations.

{\bf KKH30, NGC2146sat}.  Two isolated dwarf galaxies with  low surface brightnesses and unknown radial
velocities.  There are no structural details of them, but both galaxies have fully symmetric shapes.  $H\alpha$-fluxes have
not been detected from them, but according to the GALEX survey, both are faint $FUV$ sources.  Based on the
combination of these characteristics, they have been classified as dTr type dwarf galaxies, intermediate between dIr
and dSph.  The distances to the two galaxies are extremely uncertain.  It is possible that these dwarf systems are
peripheral satellites of the  peculiar galaxy NGC~2146, the  distance to which is estimated to be 18 Mpc [25].

{\bf AGC188955, AGC749315, AGC200499}.  Dwarf galaxies of later types observed in the ''blind'' ALFALFA
$HI$-survey [26].  Compact star formation regions can be seen in each of these three objects.

{\bf LV0853+3318, LV0926+3343}.  Two emission dwarf galaxies in the region of the nearby Gemini-Leo cosmic
void.  The first is a satellite of the  massive spiral NGC~2683, the  distance to which is ascribed [27] to its dwarf
companion.

{\bf NGC 3377, NGC 3384}.  These type E and S0 galaxies belong to the brightest members of the Leo~I group.
The distances to them have been measured using surface brightness fluctuations [28].   The central part of NGC~3377
manifests emission in the $H\alpha$-line, as well as in the far ultraviolet.

{\bf NGC 3413}. A peculiar compact galaxy of a later type with an emission ring seen nearly from the edge-on
and a bright central emission region.

{\bf UGC 7257}.  An isolated Sm galaxy with radial velocity $V_{LG} = 957$ km/s at a distance of 8.8 Mpc.  This galaxy
is distinguished by an abundance of star formation regions that form a spiral-shaped structure with a bright $HII$-complex at 
its northern edge.

{\bf UGC 7321}.  This Sd galaxy seen strictly edge-on, has radial velocity  $V_{LG} = 339$ km/s, and is at a distance
of 17.2 Mpc [2].  It belongs to the unusual Coma~I group of galaxies which has a peculiar velocity of about
--700 km/s [29].  The emission disk of this galaxy, outlined by $HII$-regions, has a record high axis ratio of  $a/b = 38$.
NGC~7321 is the closest and most distinct example of a ''superthin'' disk galaxy.

{\bf LV1219+4718, KK 136}.  Both galaxies are indicated in the data base [3] as presumed satellites of the massive
spiral NGC~4258, the periphery of which is visible in the lower right corner of the image of LV1219+4718.   The
radial velocities of these galaxies have not yet been measured.  The regular shape of the dSph structure of the galaxy
KK~136 and its lack of $H\alpha$ and FUV emission may be indirect evidence of gas being swept out  from KK~136 as it
interacts with NGC~4258.  In the case of the dIr galaxy LV1219+4718, the absence of  $H\alpha$-emission plus a substantial
FUV flux probably indicates that this is a distant background galaxy with a substantial radial velocity that shifts the
 $H\alpha$-line beyond the limits of the filter that was used.

{\bf UGC 7427}.  This irregular dwarf system has a dumbbell shape with two ring-like emission regions.  It may
be the result of the merger of two dwarf galaxies with roughly equal masses.

{\bf DDO 123 = UGC 7534}.   This is an Sm type galaxy with a bright star projected onto its southern side.  The body
of the galaxy contains many small compact $HII$ regions.  The horizontal line in the image is caused by  the passage
of a satellite during the exposure time.

{\bf SBS1224+533 = MCG+09-20-182}.  A blue compact galaxy from the Second Byurakan Survey [30] with a
radial velocity of  $V_{LG}  = 390$ km/s.

{\bf DDO 131}.  This is an irregular galaxy with a low surface brightness.  It is probably  a satellite of the spiral galaxy
NGC~4559, to which the distance is 8.1 Mpc.  Besides compact $HII$ regions, it has 4 emission rings  with angular
diameters of 5--10$^{\prime\prime}$ which may be residues of supernova outbursts.

{\bf NGC 4509 = Mrk 773}.  A compact dwarf galaxy  with a high surface brightness and a loop of small  emission knots
on its northern side.

{\bf IC 3583 = VCC 1686}.  An Sm type galaxy with a radial velocity of  $V_{LG} = 1024$ km/s.  In the sky it lies near
the center of the cluster of galaxies in Virgo. Given its distance, $D$ = 9.5 Mpc measured by the Hubble Space Telescope
[31], IC~3583 is not a member of a cluster but lies in front of a cluster at a distance of $\sim7$ Mpc from its center.  In
the continuum image, the galaxy has an asymmetric shell that gives no indication of $H\alpha$  emission.

{\bf UGCA 294 = Haro33}.  A blue compact dwarf galaxy with emission knots that are almost in contact with
one another.

{\bf IC 3840}.  An irregular dwarf galaxy of  low surface brightness.  There is a narrow emission  arc or 
a fragment of an annular structure on its northern side.

{\bf DDO 169NW}.  A diffuse ``granular'' satellite of the galaxy  DDO~169 = UGC~8331 lying 3$^{\prime}$ to the north west
of its center.  The ratio of the  hydrogen mass to the luminosity of DDO 169NW is  high [32].  The major $H\alpha$-flux
of DDO~169NW fits within a faint diffuse component distributed over the entire body of the  satellite.

{\bf CGCG 189-050}.  This is a compact blue galaxy with radial velocity  $V_{LG} = 368$ km/s.  It is probably a distant
satellite of the massive  spiral galaxy NGC~4736.

{\bf AF 7448-001}. An irregular dwarf galaxy discovered in the AGES $HI$-survey [33].  The major  $H\alpha$-flux from it
is concentrated in a single compact $HII$ region.

\section{Concluding comments.}

The data in the Table 1 show that the star formation rate has been estimated by two independent methods for
more than 3/4 of the observed galaxies: based on the flux in the $H\alpha$-line and based on the far ultraviolet flux $FUV$.
The average difference in the star formation rates according to these two methods is
$\langle\log SFR(H\alpha) - \log SFR(FUV)\rangle = -0.18\pm0.10$ 
   with a mean square deviation of 0.40.  It has shown repeatedly [34 ,12, 13]
that determinations of the star formation rate based on the  $H\alpha$-flux for dwarf galaxies yield a systematic underestimate
of the $SFR$ compared to measurements based on the $FUV$ flux.  This difference is larger when  the luminosity of the
dwarf galaxy is lower and its surface brightness is lower.  It has been noted [34]  that the initial stellar mass function
can be substantially different for dwarf and spiral galaxies at its bright end.  Because of this, the empirical normalization
  $ SFR(H\alpha) = SFR(FUV)$  for massive disks is not entirely suitable for dwarf systems.  Individual differences in
the estimates of $SFR(H\alpha)$ and $SFR(FUV)$ can also be caused by the variability in the star formation rate that is typical
of dwarf galaxies.  The $SFR(FUV)$ estimates typically apply to a  time interval of $\sim$100 Myr, while the values of
$SFR(H\alpha)$ correspond to an interval of  $\Delta t\sim 10$ Myr. At the time of the maximum of a star formation burst,
 $SFR (H\alpha)/SFR(FUV)$  can be substantially greater than unity for irregular and blue compact galaxies.

According to observational data [12.35,36], the specific star formation rate  $sSFR = SFR/M^*$ relative to unit
stellar mass of the galaxy does not exceed an  upper bound of $(SFR)_{max} \simeq -9.4\cdot$ dex (yr$^{-1}$). 
 This empirical condition
is satisfied for 99\% of galaxies of all morphological types and is an important characteristic of the process by which
gas is transformed into stars in the current epoch.  As the data in the last columns of the table show, this limit is
obeyed by the estimates of $SFR(H\alpha)$  and $SFR(FUV)$ for all the galaxies we have observed.  We emphasize that the
existence of an empirical upper bound $(SFR)$ max for galaxies has not yet received a clear theoretical explanation.

{\bf Acknowledgements}. This work was supported by grants No. 15--52--45004 and 13--02--00780 of the Russian Foundation for Basic
Research (RFFI) and the new observational data for completing the data base at \\ http://www.sao.ru/lv/lvgdb were
obtained with the support of a grant from the Russian Science Foundation (project No. 14-12-00965).

\bigskip

{\bf REFERENCES.}

1. {\em I.D.Karachentsev, V.E.Karachentseva, W.K.Huchtmeier, and D.I.Makarov}, Astron. J., {\bf 127}, 2031, 2004 =CNG

 2. {\em I.D.Karachentsev, D.I.Makarov, and E.I.Kaisina}, Astron. J., {\bf 145}, 101, 2013 =UNGC
 
 3. {\em E.I.Kaisina, D.I Makarov, I.D.Karachentsev, and S.S.Kaisin}, Astrophys. Bull., {\bf 67}, 115, 2012
 
 4. {\em S.S.Kaisin, and I.D.Karachentsev}, Astrophysics, {\bf 49}, 287, 2006
 
 5. {\em I.D.Karachentsev, and S.S.Kaisin}, Astron. J., {\bf 133}, 1883, 2007
 
 6. {\em S.S.Kaisin, and I.D.Karachentsev}, Astron. and Astrophys., {\bf 479}, 60, 2008
 
 7. {\em I.D.Karachentsev, and S.S.Kaisin}, Astron. J., {\bf 140}, 1241, 2010
 
 8. {\em S.S.Kaisin, I.D.Karachentsev, and E.I.Kaisina}, Astrophysics, {\bf 54}, 315, 2011
 
 9. {\em S.S.Kaisin, and I.D.Karachentsev}, Astrophysics, {\bf 56}, 305, 2013
 
10. {\em S.S.Kaisin, and I.D.Karachentsev}, Astrophys. Bull., {\bf 68}, 381, 2013

11. {\em S S. Kaisin, and I.D. Karachentsev},  Astrophys. Bull., {\bf 69}, 390, 2014

12. {\em I.D. Karachentsev, and E.I.Kaisina}, Astron.J., {\bf 146}, 46, 2013

13. {\em J.C. Lee, R. C. Kennicutt, J. G. Funes, et al.}, Astrophys. J. , {\bf 692}, 1305, 2009

14. {\em A.Gil de Paz, S.Boissier,  B.F.Madore  et al.},  Astrophys.J.Suppl., {\bf 173}, 185, 2007

15. {\em A Begum, J.N.Chengalur, I.D.Karachentsev , M.E. Sharina, and S.S.Kaisin}, MNRAS, {\bf 386}, 1667, 2008

16. {\em S.Roychowdhury, J.N. Chengalur, A. Begum, and I.D. Karachentsev}, MNRAS, {\bf 404L}, 60, 2010

17. {\em A.V. Moiseev}, Astrophys. Bull.,  {\bf 69}, 1, 2014

18. {\em A.V. Moiseev, A.V. Tikhonov, and A. Klypin}, MNRAS, {\bf 449}, 3568, 2015

19. {\em D.J. Schlegel, D.P. Finkbeiner, and M.Davis}, Astrophys. J., {\bf 500}, 525, 1998

20. {\em R.C. Kennicutt}, Annu. Rev. Astronom. Astrophys. {\bf 36}, 189, 1998

21. {\em M.A.W. Verheijen},  Astrophys. J., {\bf 563}, 694, 2001

22. {\em A.P. Huxor, A.D.Mackey, A.M.N.Ferguson, et al.}, MNRAS, {\bf 442}, 2165, 2014

23. {\em N. Trentham, and R.B. Bully}, MNRAS, {\bf 398}, 722, 2009

24. {\em I.D. Karachentsev, M.E. Sharina, D.I. Makarov, et al.}, Astrofisics, {\bf 58}, 331, 2015

25. {\em A. Adamo, L.J. Smith, J.S. Gallagher, et al.}, MNRAS, {\bf 426}, 1185, 2012
 
26. {\em M.P. Haynes, R. Giovanelli, A.M. Martin, et al.}, Astron. J., {\bf 142}, 170, 2011

27. {\em I.D. Karachentsev, R.B. Tully, L.N.Makarova, et al.},  Astrophys. J., {\bf 805}, 144, 2015

28. {\em L.Ferrarese, J.R.Mould, R.C.Kennicutt, et al.},  Astrophys. J., {\bf 529}, 745, 2000

29. {\em I.D. Karachentsev, O.G. Nasonova, and H.M. Courtois}, Astrophys. J., {\bf 743}, 123, 2011

30. {\em B.E. Markarian}, Astrofizics, {\bf 12}, 389, 1976

31. {\em I.D. Karachentsev, R.B. Tully, Po-Feng Wu, E.J. Shaya, and A.E. Dolphin}, Astrophys. J., {\bf 782}, 4, 2014

32. {\em R.A. Swaters}, PhD Thesis, Groningen, 1999

33. {\em R.Taylor, R.F.Minchin, H.Herbst, and R.Smith}, MNRAS, {\bf 442L}, 46, 2014

34. {\em J.Pflamm-Altenburg, C.Weidner, and P.Kroupa}, Astrophys. J., {\bf 671}, 1550, 2007

35. {\em I.D.Karachentsev, V.E.Karachentseva, O.V.Melnyk, and H.M.Courtois}, Astrophys. Bull., {\bf 68}, 243, 2013

36. {\em V.E.Karachentseva, O.V.Melnyk, and I.D.Karachentsev}, Astrofizics, {\bf 57}, 5, 2014

\clearpage
\topmargin=-1cm
 \begin{figure}
\includegraphics[scale=1.0]{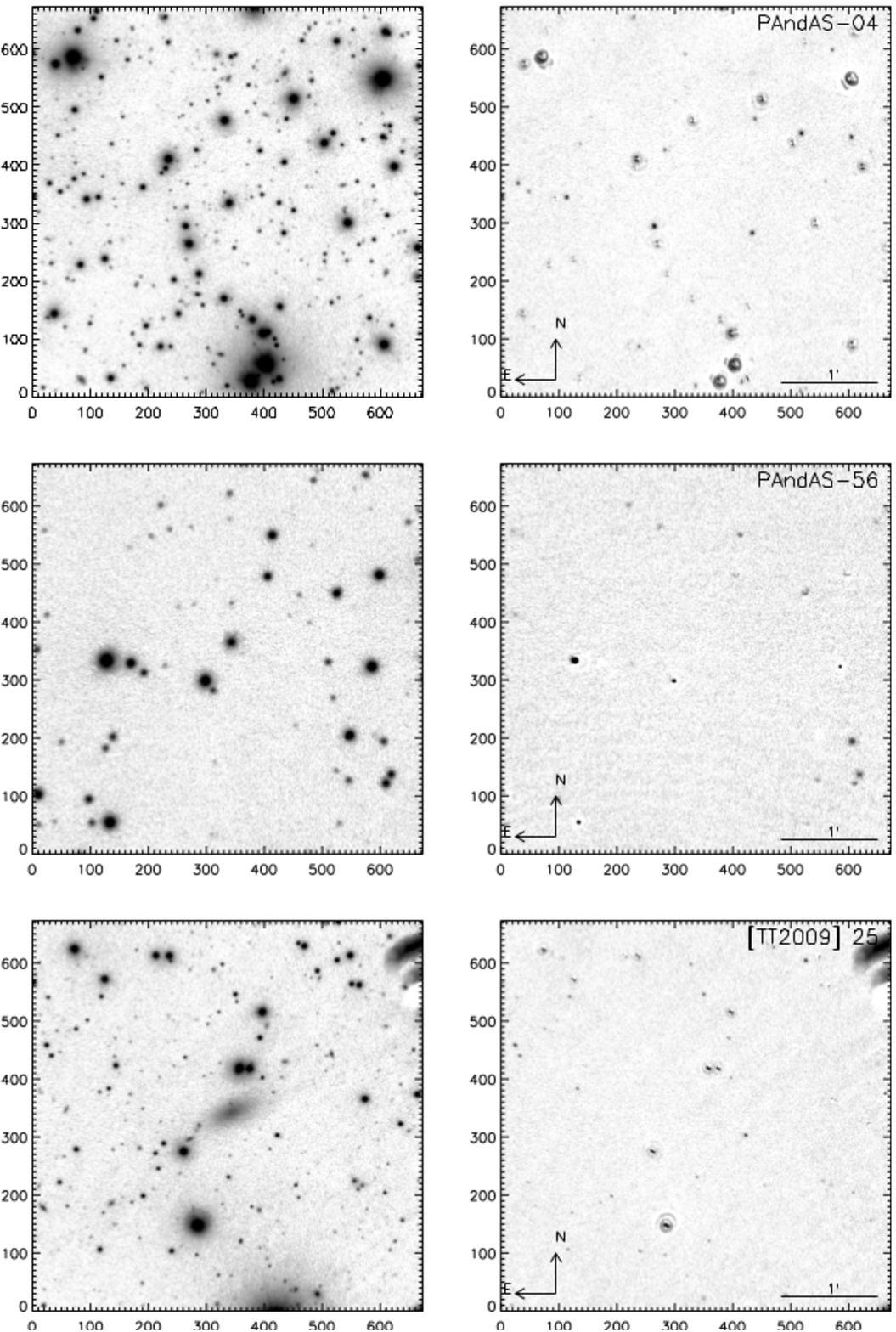}
\caption{A mosaic of images of objects in the Local volume.  The
images on the left of each pair are the sum of exposures in the $H\alpha$  line
and the continuum, and the images on the right correspond to the
``$H\alpha$-continuum'' difference.  The angular scale and orientation are
indicated in the bottom corners of the images on the right.}
\end{figure}

\setcounter{figure}{0}
 \begin{figure}
\includegraphics[scale=1.0]{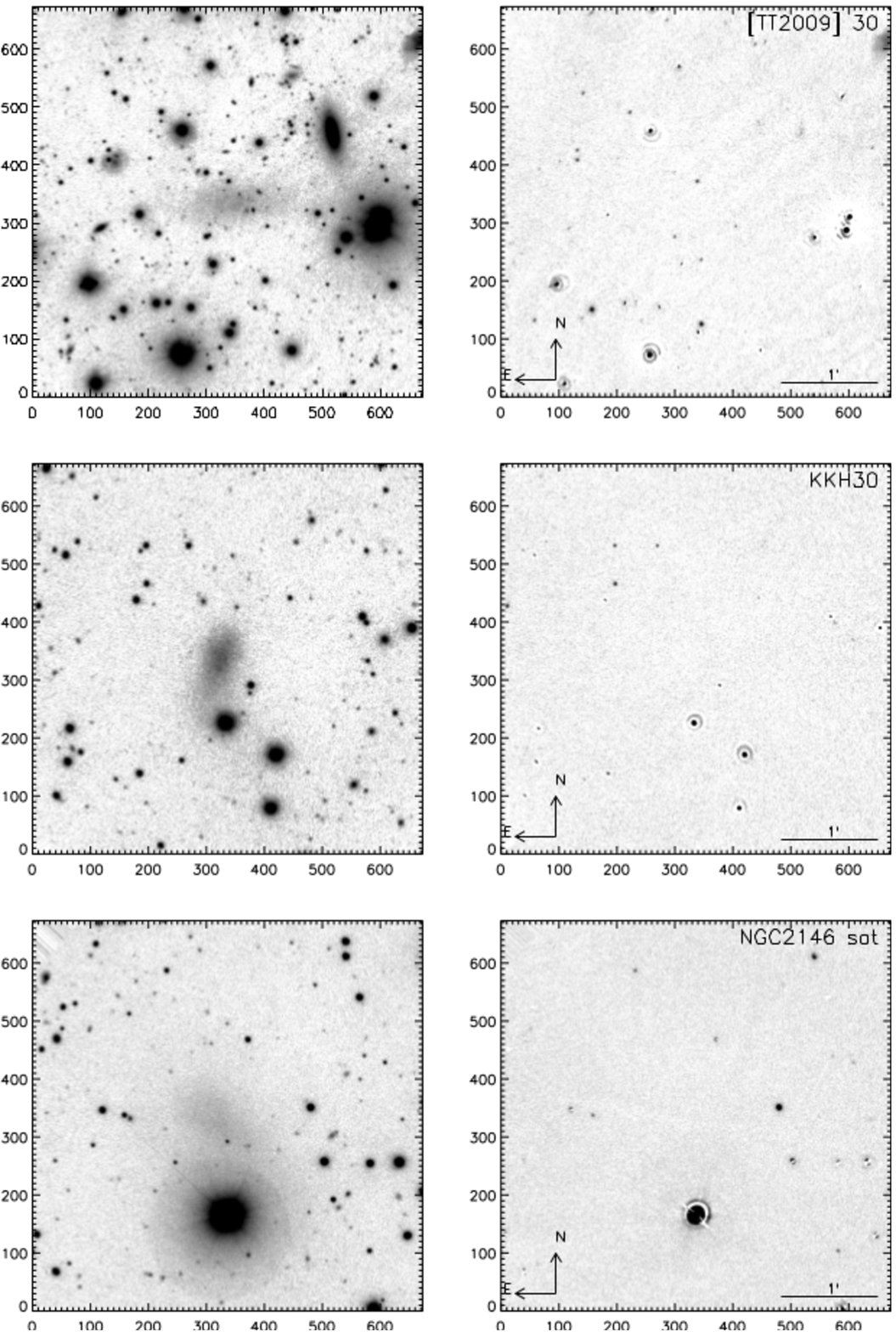}  
\caption{Continued.}
\end{figure}
\setcounter{figure}{0}
\begin{figure}
\includegraphics[scale=1.0]{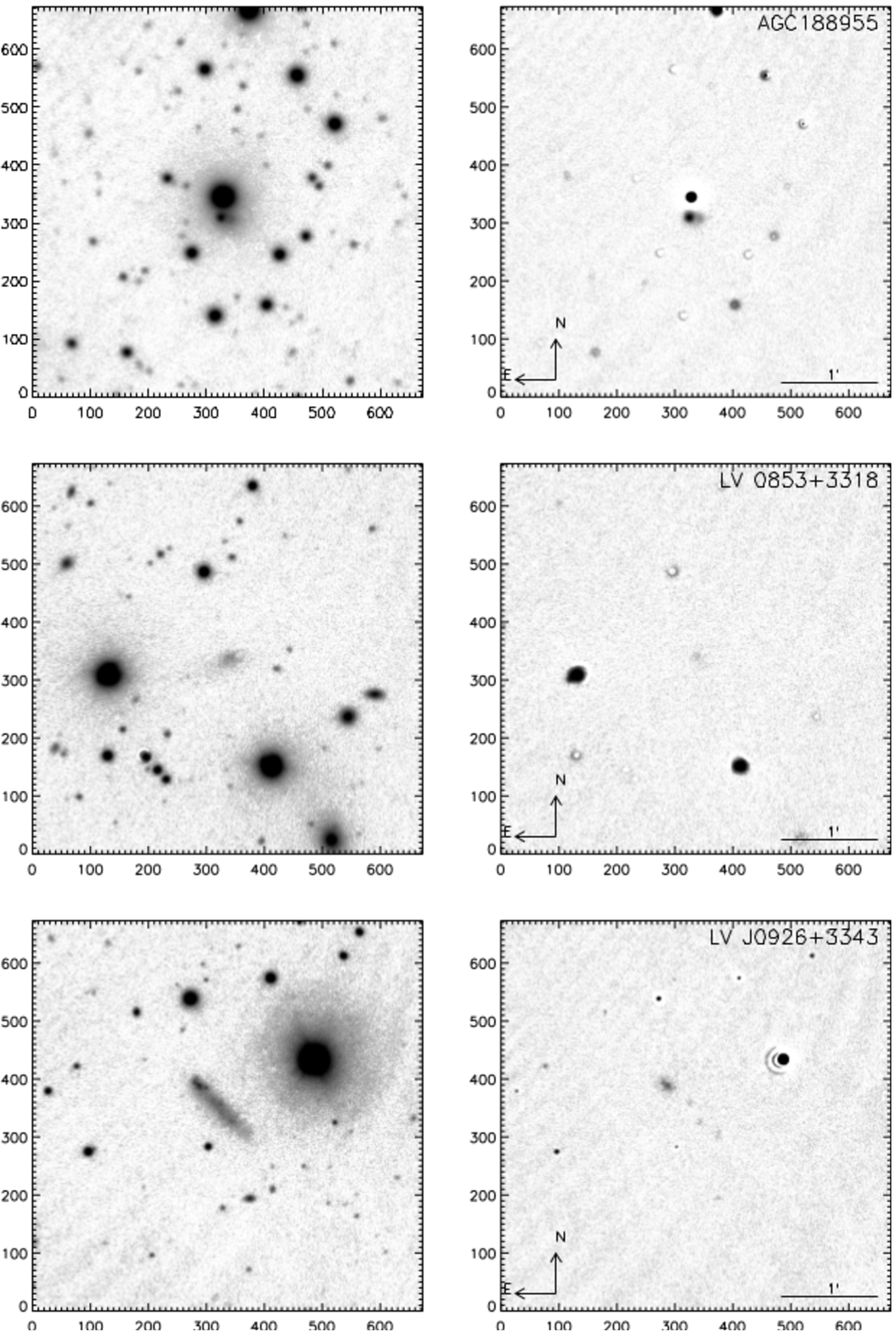}  
\caption{Continued.}
\end{figure}

\setcounter{figure}{0}
\begin{figure}
\includegraphics[scale=1.0]{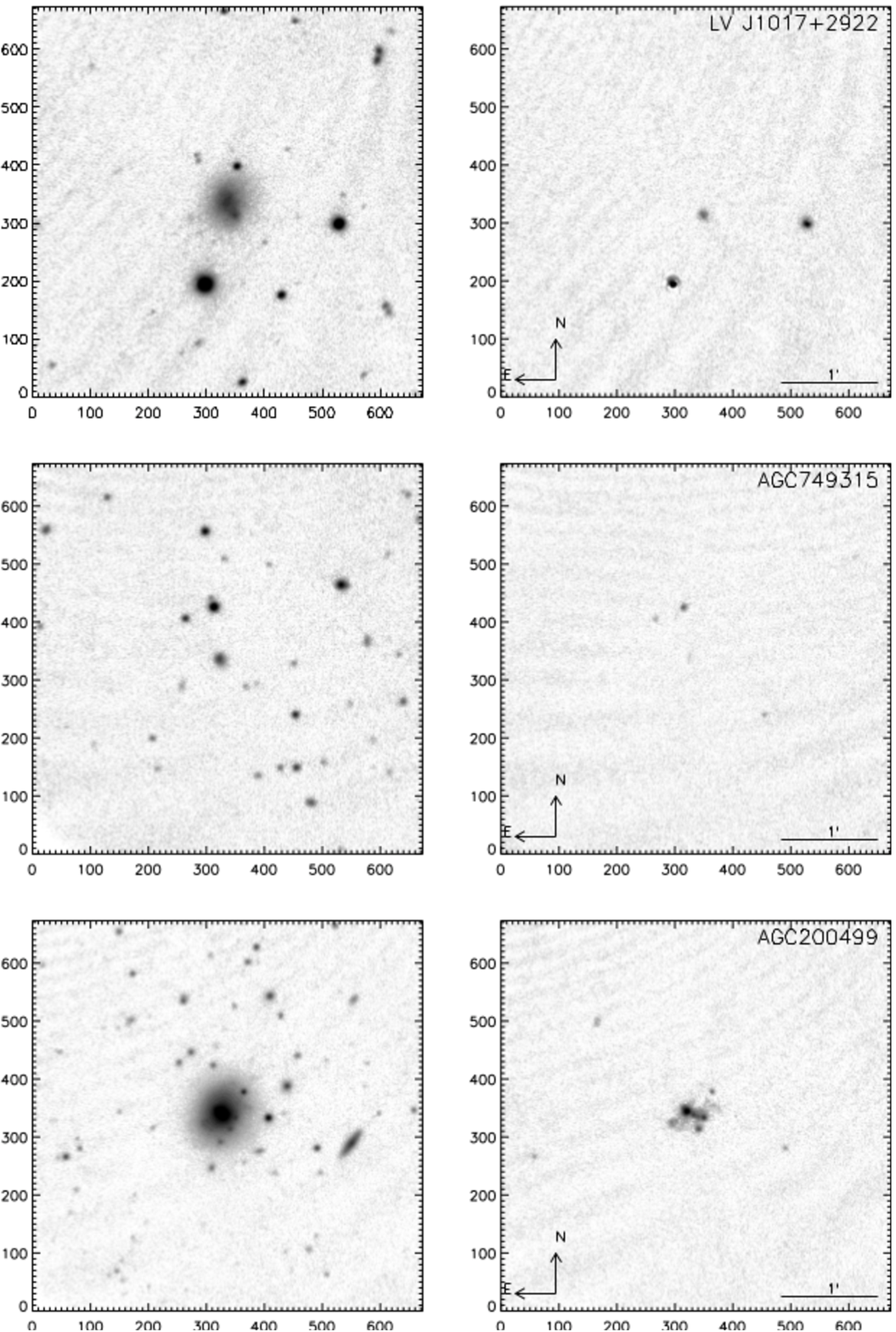}  
\caption{Continued.}
\end{figure}
\setcounter{figure}{0}
\begin{figure}
\includegraphics[scale=1.0]{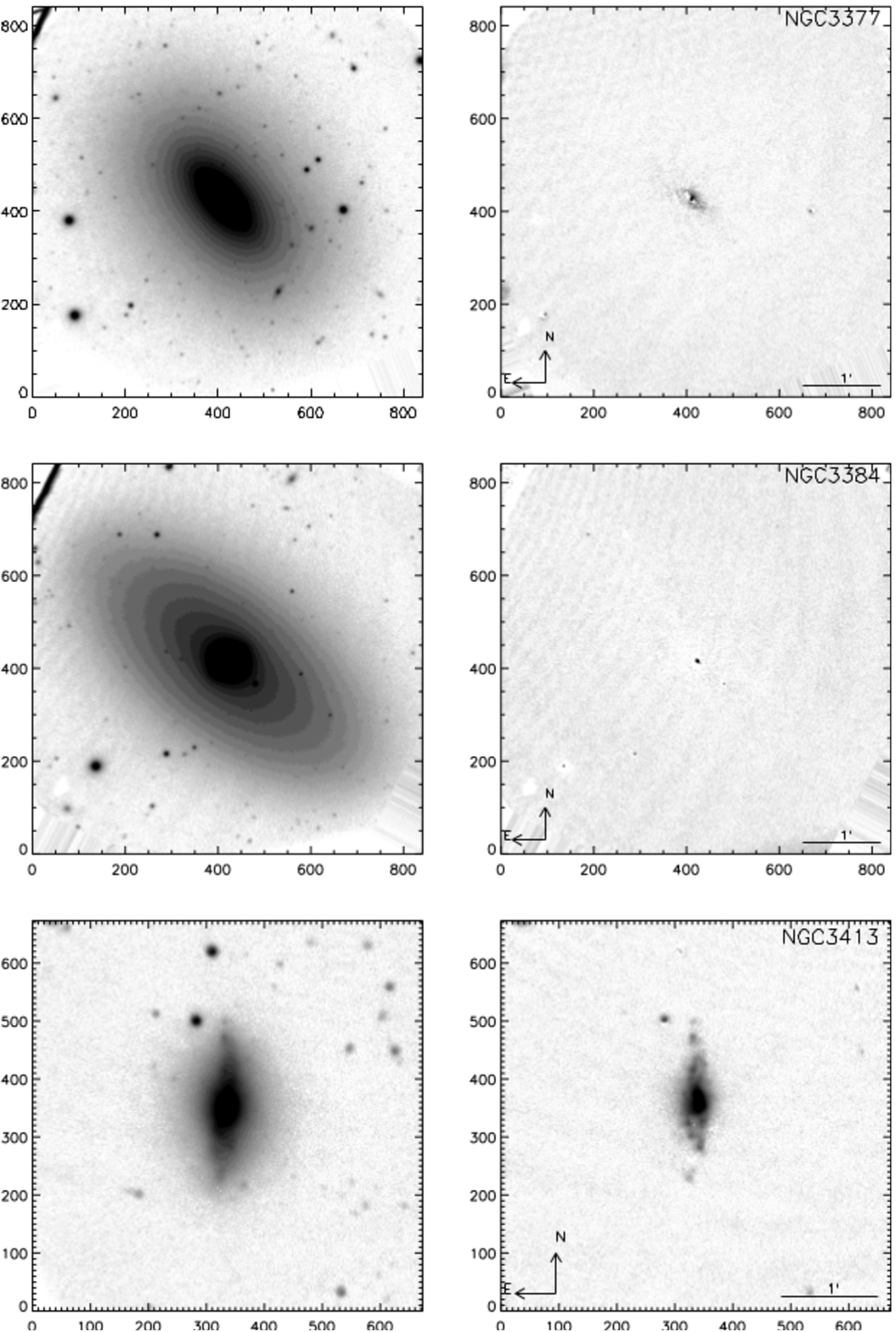}  
\caption{Continued.}
\end{figure}
\setcounter{figure}{0}
\begin{figure}
\includegraphics[scale=1.0]{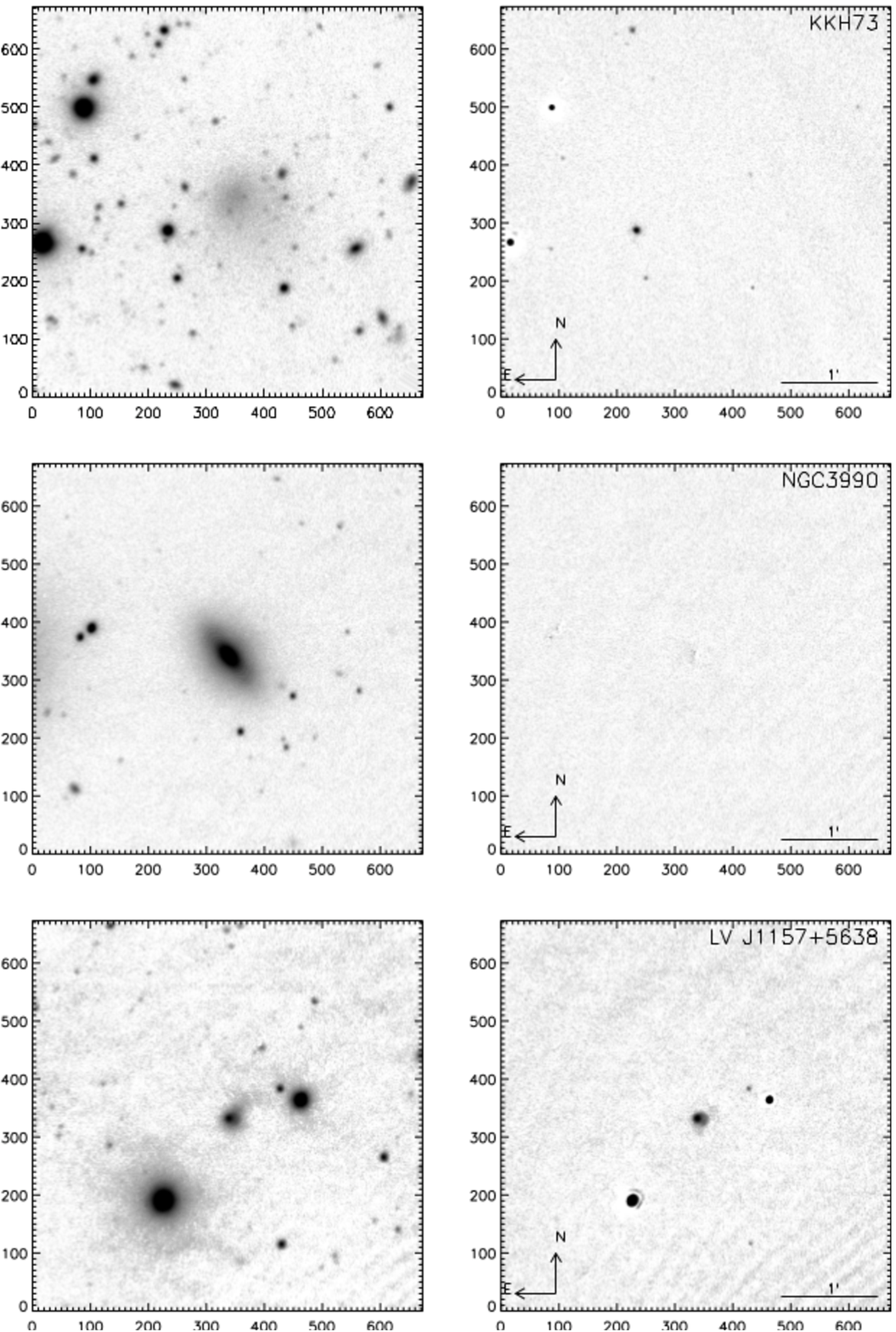}  
\caption{Continued.}
\end{figure}
\setcounter{figure}{0}
\begin{figure}
\includegraphics[scale=1.0]{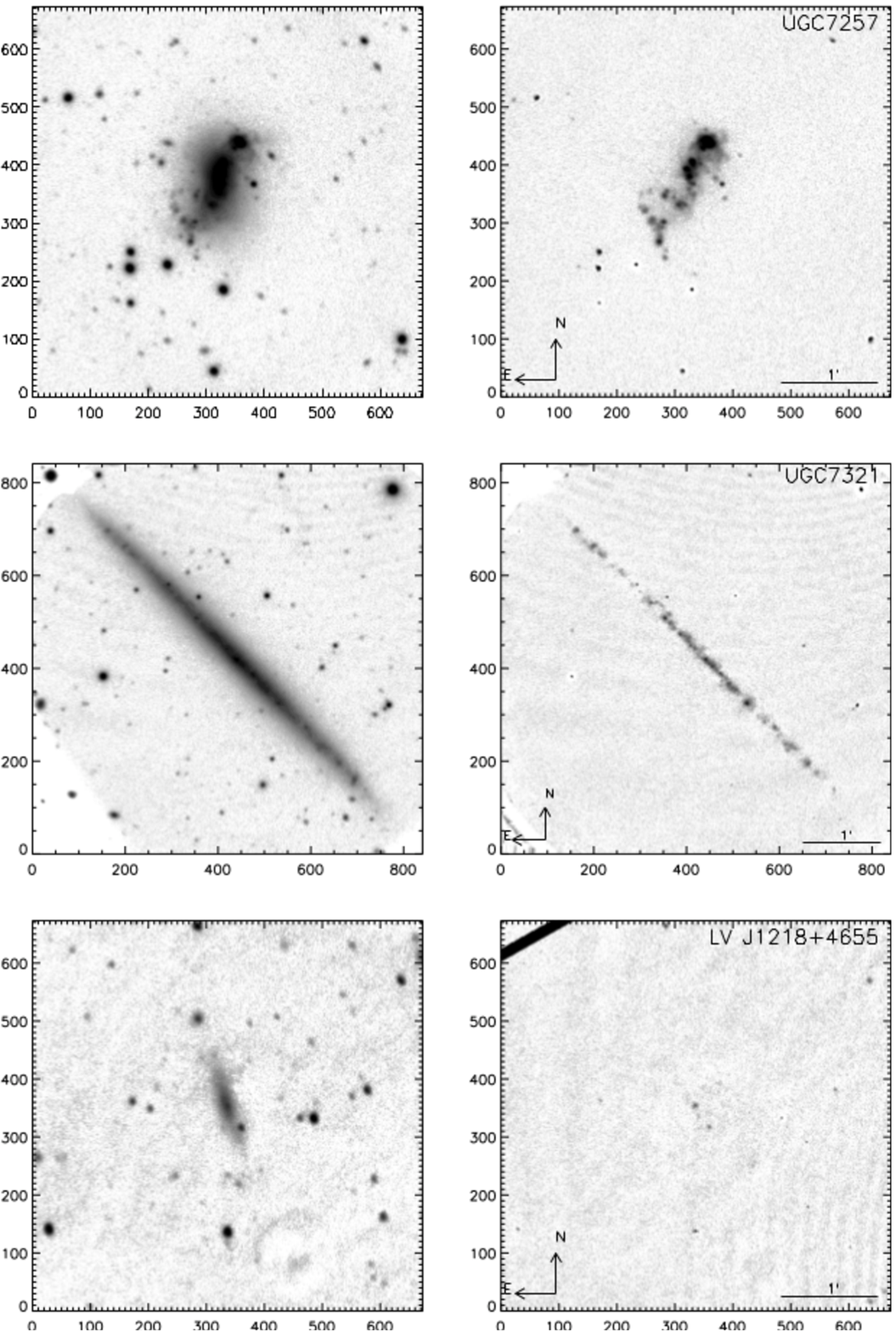}  
\caption{Continued.}
\end{figure}
\setcounter{figure}{0}
\begin{figure}
\includegraphics[scale=1.0]{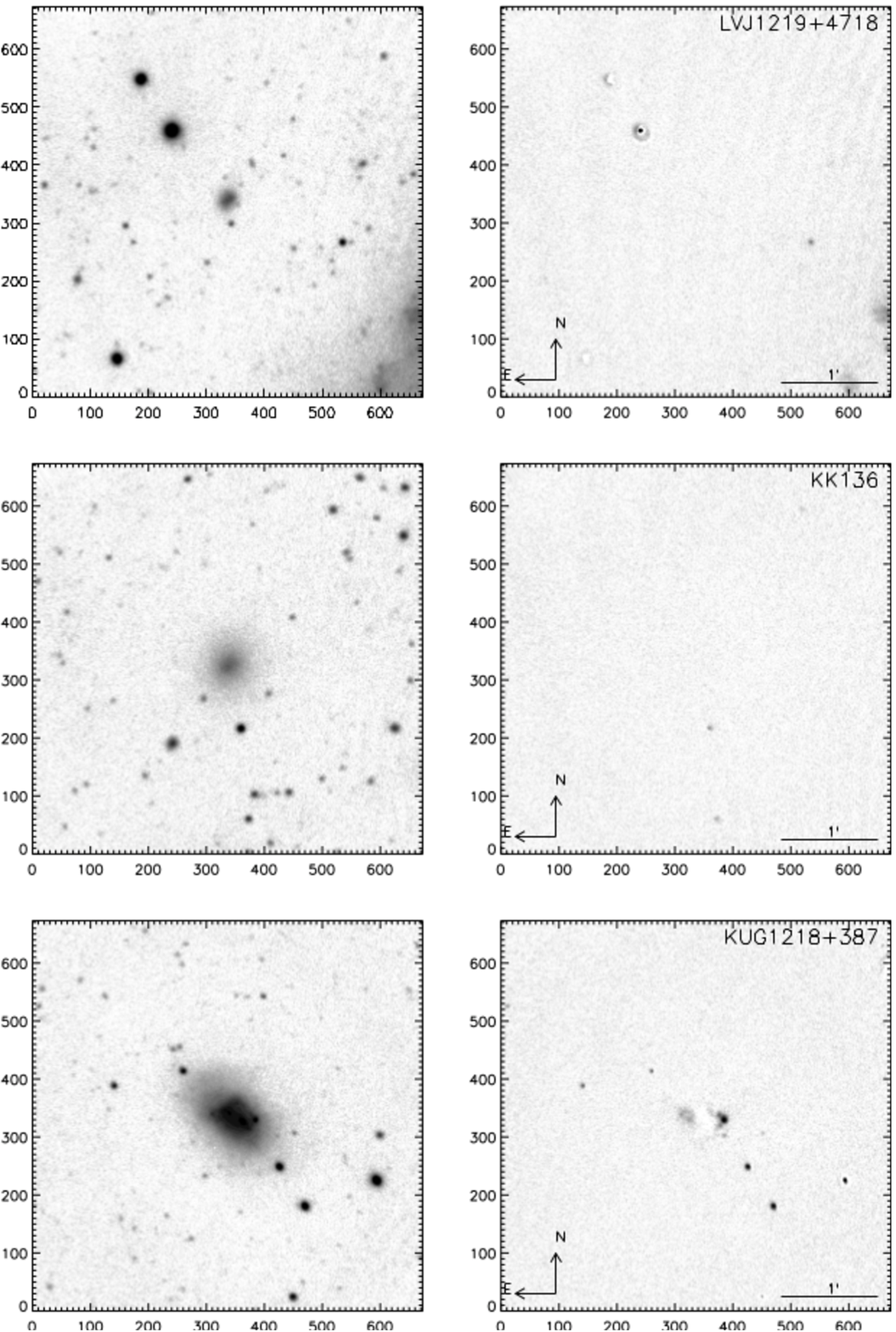}  
\caption{Continued.}
\end{figure}
\setcounter{figure}{0}
 \begin{figure}
\includegraphics[scale=1.0]{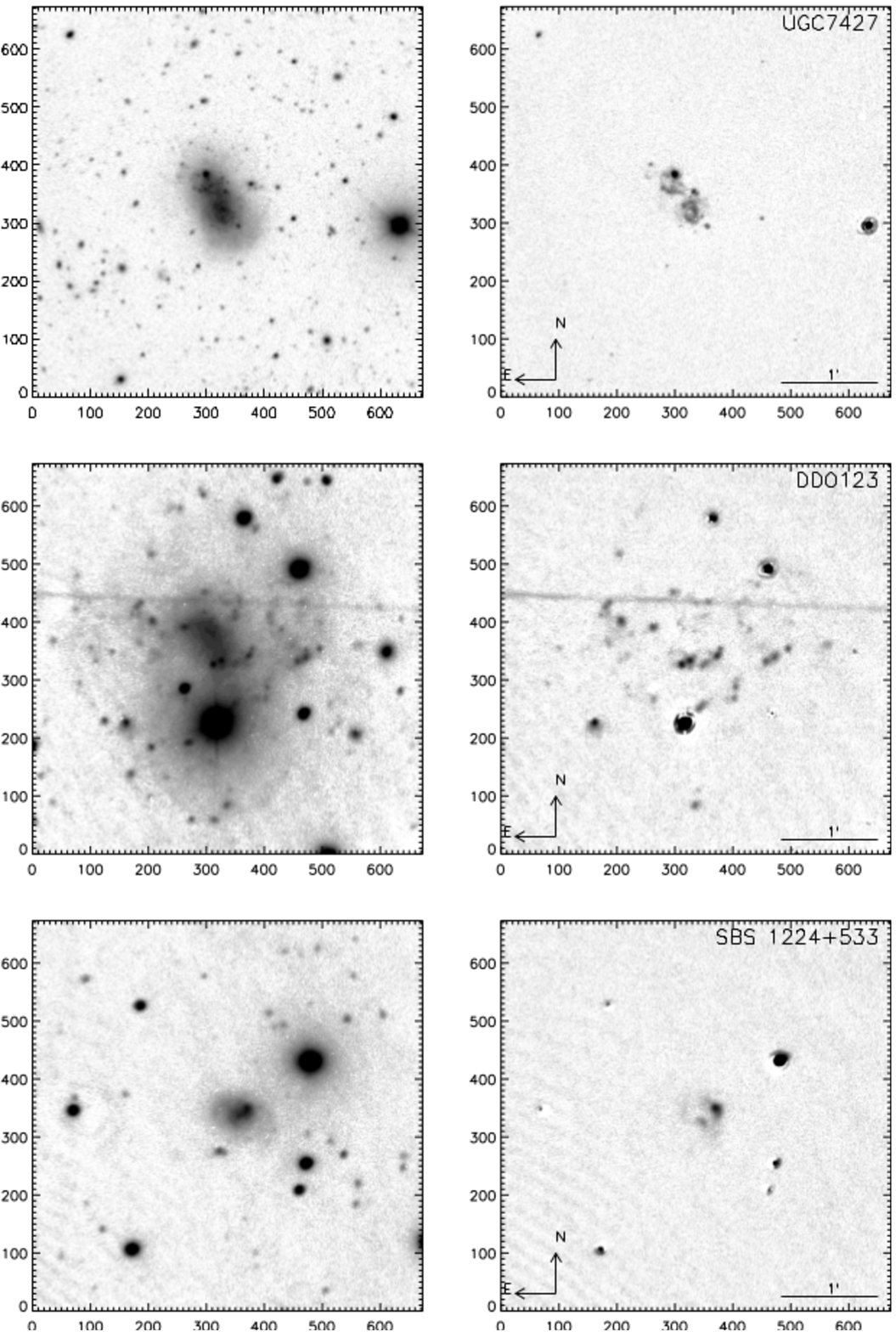}  
\caption{Continued.}
\end{figure}
 \setcounter{figure}{0}
 \begin{figure}
\includegraphics[scale=1.0]{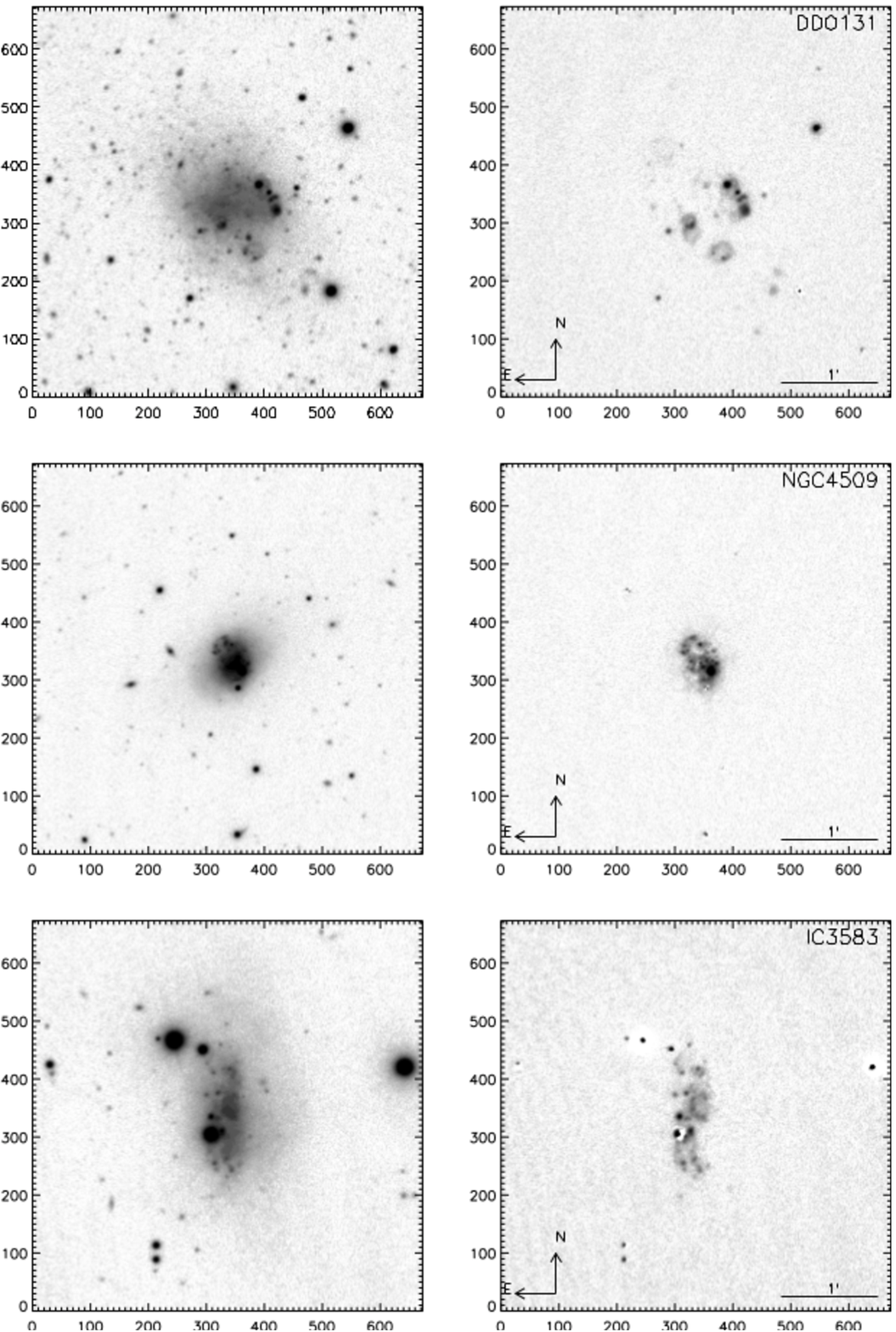}  
\caption{Continued.}
\end{figure}
\setcounter{figure}{0}
\begin{figure}
\includegraphics[scale=1.0]{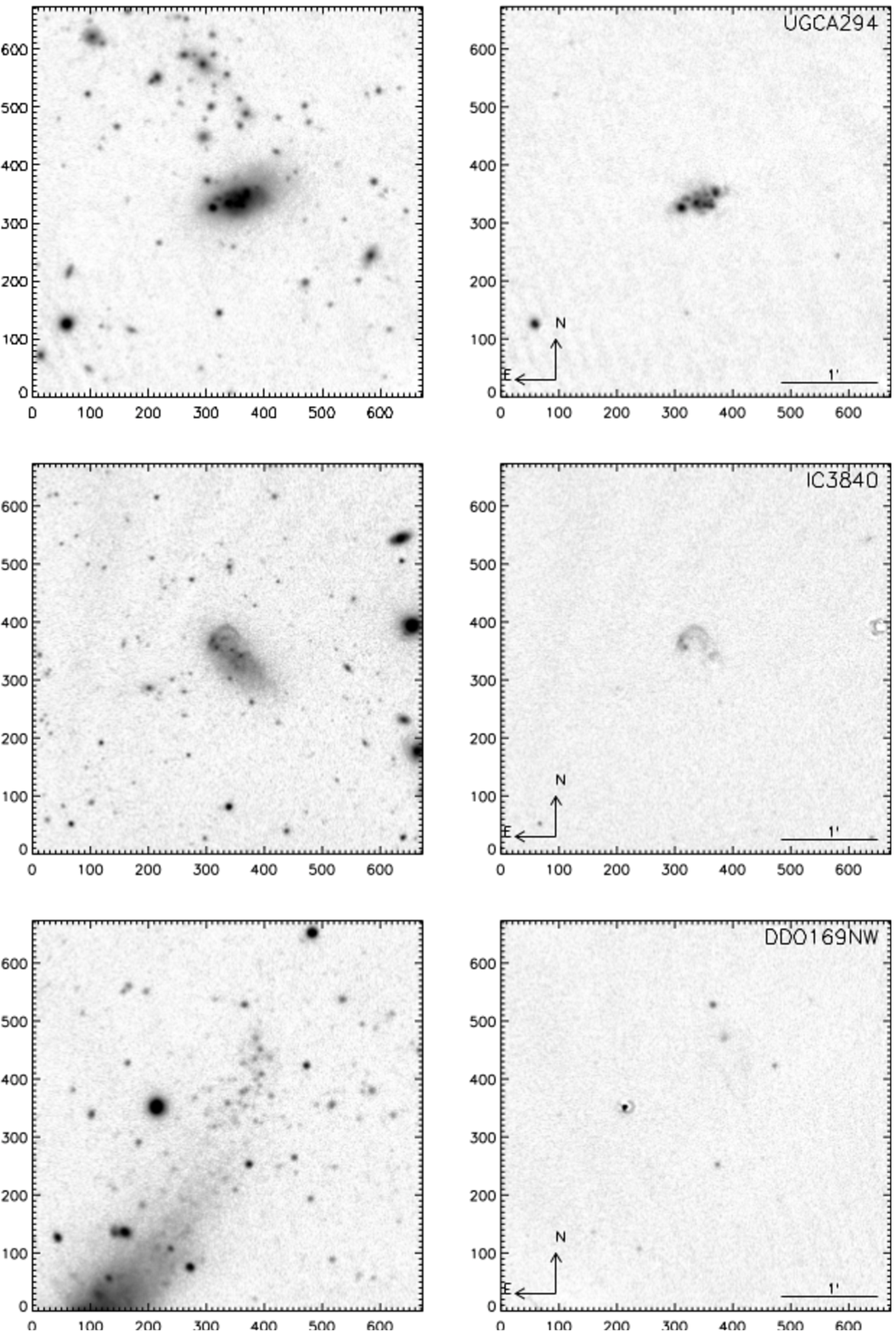}  
\caption{Continued.}
\end{figure}
\setcounter{figure}{0}
\begin{figure}
\includegraphics[scale=1.0]{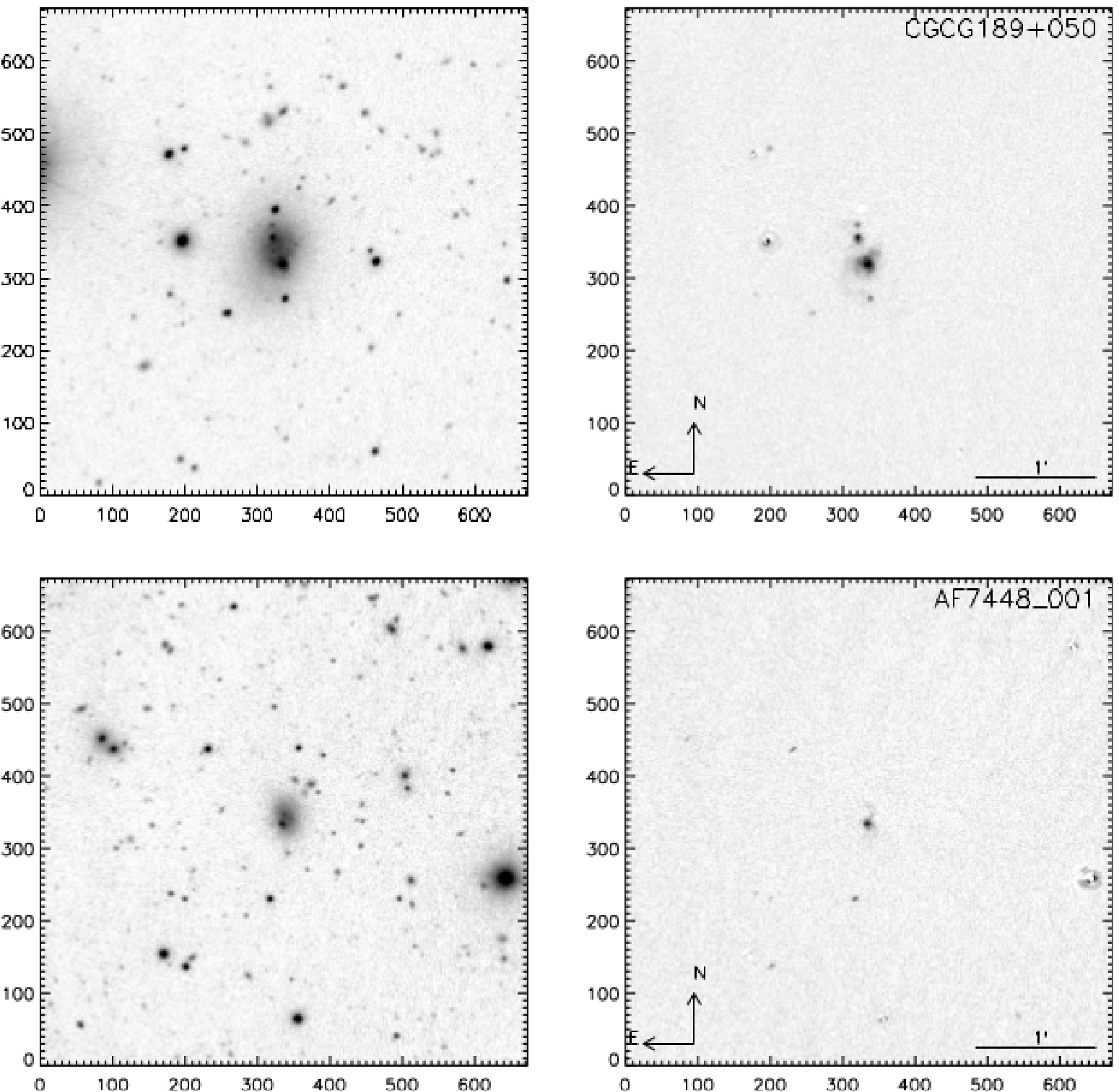}  
\caption{Conclusion.}
\end{figure}

\end{document}